\documentclass[conference]{IEEEtran}

\usepackage{cite}
\usepackage[pdftex]{graphicx}
\usepackage[cmex10]{amsmath}
\usepackage{algorithm}
\usepackage{algorithmic}
\usepackage{multirow}

\usepackage[caption=false,font=footnotesize]{subfig}

\newcommand{\ket}[1]{\left|#1\right\rangle}

\begin{document}

\title{Programming a Topological Quantum Computer}

\author{\IEEEauthorblockN{Simon Devitt$^1$
    \qquad Kae Nemoto$^1$}\\[0.1cm]
  \IEEEauthorblockA{
  \begin{tabular}{c@{\hspace{0cm}}c}
    $^1$National Institute of Informatics \\
    2-1-2 Hitotsubashi, Chiyoda-ku \\
   Tokyo, Japan \\
   \{devitt$|$nemoto\}@nii.ac.jp
  \end{tabular}
}}
\maketitle
\begin{abstract}
Topological quantum computing has recently proven itself to be a powerful computational model when 
constructing viable architectures for large scale computation.  The topological model is constructed from 
the foundation of a error correction code, required to correct for inevitable hardware faults that 
will exist for a large scale quantum device.  It is also a measurement based model of quantum computation, 
meaning that the quantum hardware is responsible \emph{only} for the construction of a large, 
computationally universal quantum state.  
This quantum state is then strategically consumed, allowing for the realisation of 
a fully error corrected quantum algorithm.  The number of physical qubits needed by the 
quantum hardware and the amount of time required to implement an algorithm is dictated by the manner in 
which this universal quantum state is consumed.  In this paper we examine the problem of algorithmic 
optimisation in the topological lattice and introduce the required elements that will be needed when 
designing a classical software package to compile and implement a large scale algorithm on a 
topological quantum computer.
\end{abstract}

\section{Introduction}

Quantum information science has been one of the extraordinary success stories of theoretical and experimental physics in the last 20 years.  Not only has a complete theoretical framework for universal quantum computation 
been established, but algorithms have also been discovered that can vastly outperform their classical counterparts \cite{S94}.  Multiple physical systems can now routinely demonstrate fabrication and control of 
small arrays of quantum bits (qubits) \cite{LJLNMO10}. The progress of experimental systems has allowed, in 
recent years, for the development of multiple quantum architectures demonstrating how a large-scale multi-million qubit machine could be built \cite{MTCCC05, FTYSPW07, DFSG08, MLFY10, JMFMKLY10, YJG10}. 

Even with the extraordinary level of experimental control at the quantum level, imperfections in qubit manufacturing 
and control still lead to errors in quantum logic operations, currently at the level of a few percent (for even the best systems).  This 
level of hardware error is unacceptable for large scale algorithms and it is unlikely that hardware imperfections 
can be reduced to an acceptable level anytime in the near future.  This problem was well known since the 
first development of quantum information science and methods for achieving large scale computation using 
inherently faulty components was quickly formulated \cite{DMN09}.  Borrowing from classical information science, Quantum 
Error Correction (QEC) and Fault-Tolerant Quantum Computation (FTQC) allowed for arbitrarily large 
algorithms to be run with faulty components, provided that the error associated with each component 
is below a certain \emph{threshold} level \cite{DMN09}.  

Theoretical development of QEC and FTQC in the past ten years has been focused on the construction of 
codes that are amenable to physical hardware designs and increasing the fault-tolerant threshold to a 
level achievable by experiment in the next decade. 
The topological model of QEC \cite{K03,DKLP02,RHG07} has shown itself to be more promising compared with  many other 
long-standing techniques and currently forms the basis of effectively all modern quantum-computing architectures \cite{DFSG08, MLFY10, JMFMKLY10, YJG10}.  Each of these hardware designs utilise a different physical 
system that define the qubit and all allow for a broad range of physical operational speeds, physical component 
sizes and associated ancillary technology such as cryogenic cooling and ultrahigh vacuums.  However, 
architectures based on the topological model all have one thing in common; namely that the realisation of 
a large algorithm is essentially independent of the quantum hardware.  

This method of computation is very abstract when compared to classical computer science.  One of the more bizarre aspects of this model is that  the physical hardware \textit{doesn't actually perform any real computation}. Instead, the hardware is only responsible for producing a very large three-dimensional lattice of qubits which are all linked together to form a single, massive, universal quantum state. This quantum state forms the \textit{workbench} of the computation and information is created, processed and read-out via the strategic manipulation of this massive quantum state \cite{RHG07,FG08}.  For example, if single photons are used to prepare the lattice, the entangled state can travel far from the physical location of the actual computer before each 
photon is measured and data processing begins.  The algorithmic implementation is consequently dependent 
on how this 3D lattice is consumed, rather than how it is prepared.  

How large scale algorithms relate to the total number of devices in the computer and the total amount of 
time needed for computation is ultimately related to the 3D size of the lattice that is required.  Computation 
in this model is realised via geometric shapes, known as defects, which are created and manipulated 
within the lattice.  Each pair of these defects represents a logically encoded qubit, and occupies a certain amount 
of space within the lattice.  Therefore, a 3D lattice must be prepared which is physically large enough to 
encapsulate all the defect qubits needed for the algorithm and associated logic gates.  Compilation and 
optimisation in this model requires the translation of the quantum circuit into the geometric arrangements 
of defects and a method of compactification which allows us to utilise as much of the lattice as possible, 
minimising the volume of the lattice and consequently minimising the physical resources of the computer. 

In this paper we introduce the problem of programming a topological computer and attempt to outline 
the issues required when converting and optimising an abstract quantum algorithm into the physical 
operations performed by the computer.  We will attempt to explain these concepts in a way that requires 
little knowledge of the background physics of quantum computation.  In previous work we have attempted 
to formulate a framework of algorithmic optimisation in the topological model \cite{PDNP12}, however in this 
paper we will restrict ourselves to introducing the nature of the classical problem, rather than discussing 
possible solutions. 

\section{Background}

\subsection{Quantum Computing}

Quantum computing can be, to a certain extent, described by building parallels to classical computing and a comprehensive 
review of quantum information and computation can be found in Ref. \cite{NC00}. The concept of classical bit has its quantum counterpart, called a quantum bit (\emph{qubit}).   The binary states of a qubit ($\ket{0}$, $\ket{1}$) 
can be, for example, the polarisation state of a single photon, the spin state of a single electron or the 
direction of current flow around a loop of superconducting wire.  Unlike classical bits, a qubit can exist is 
a general superposition of the two basis states.  This quantum state, $\ket{\psi}$,  can 
can be represented as a vector $\ket{\psi}=\alpha_0\ket{0} + \alpha_1\ket{1}$, where $\alpha_0$ and $\alpha_1$ are complex numbers (called amplitudes) that satisfy a normalisation condition $|\alpha_0|^2 + |\alpha_1|^2 = 1$.  While the state of the 
qubit can be in a generalised superposition, when its state is measured, it will collapse to one of the 
two binary states, with a probability associated with the complex amplitude.
Reading the value of a bit has a quantum counterpart; measurement. Unlike classical readout, quantum measurement allows us to read out qubits in multiple ways.  The standard measurement in quantum computation, referred to as a $Z$-basis measurement.  This measurement discriminates if the qubit is in the $\ket{0}$ or $\ket{1}$ state, collapsing the wave function (terms in the superposition inconsistent with the measurement result are discontinuously 
removed) describing the qubit array.  For example, the state $\alpha_0\ket{0}+\alpha_1\ket{1}$ has a probability of $|\alpha_0|^2$ of being measured in the $\ket{0}$ state and a probability of $|\alpha_1|^2$ of being measured in the $\ket{1}$ state.  Another type of measurement is an $X$-basis 
measurement, which measures if the qubit is in the $\frac{1}{\sqrt{2}}\left(\ket{0}+\ket{1}\right)$ state or the $\frac{1}{\sqrt{2}}\left(\ket{0}-\ket{1}\right)$ state.  This type of measurement is valid as these two states are orthogonal (the wavefunctions describing these 
states have zero overlap).

A quantum gate manipulating $m$ qubits is described by a $2^m \times 2^m$ unitary matrix acting 
on a column vector of length $2^{m}$ with entries $\{\alpha_i\}$ where $\sum_{i=0}^{2^m-1}|\alpha_i|^2 = 1$ 
for $i \in \{0,..,2^{m}-1\}$ representing 
the $m$-qubit state $\ket{\phi} = \sum_{i=0}^{m-1}\alpha_i\ket{i}$.  For example, the \emph{controlled-not} gate (CNOT) acting on two qubits is defined by the following $4 \times 4$ matrix:

\begin{equation}
  \label{eq:cnot}
  \textstyle CNOT = \left(
    \begin{array}{cccc}
      1 & 0 & 0 & 0 \\
      0 & 1 & 0 & 0 \\
      0 & 0 & 0 & 1 \\
      0 & 0 & 1 & 0
    \end{array}
\right).
\end{equation}

This gate will take a general state of two qubits, $\ket{\phi} = \alpha\ket{00} + \beta\ket{01} + \gamma\ket{10} 
+ \delta\ket{11}$ and flips the state of the second qubit, conditional on the first qubit being in the 
$\ket{1}$ state.  Hence, $\text{CNOT}\ket{\phi} = \alpha\ket{00} + \beta\ket{01}+\gamma\ket{11} + \delta\ket{10}$.  

\subsection{Topological Quantum Computing}

There are two components necessary for a realistic model of quantum computation.  The first is a concept 
known as universality.  

\subsubsection{Universality} A general quantum algorithm operating on an array 
of $m$ qubits can be described as a series of $2^m \times 2^m$ unitary operations, interspersed with 
selective qubit measurement.  However, a large programmable unitary operation is unrealistic 
given the restrictions of the physical hardware.  Instead, unitaries must be decomposed into a small 
discrete set of gates (ideally operating on very few qubits) that can be combined to construct any 
desired $m$-qubit unitary.  This concept of a universal gate set was first established in the 1980's by 
Deutsch \cite{D85,D89}.  One such set of gates consists of the two qubit CNOT gate (illustrated above) and the 
following three single qubit gates,
\begin{equation}
H = \frac{1}{\sqrt{2}}
\begin{pmatrix} 1 & 1 \\ 1 & -1 \end{pmatrix} \quad P = \begin{pmatrix} 1 & 0 \\ 0 & i \end{pmatrix} \quad 
T = \begin{pmatrix} 1 & 0 \\ 0 & e^{i\frac{\pi}{4}} \end{pmatrix}
\label{Eq:Tstate}
\end{equation}
The $H$ (Hadamard) gate, takes the state $\ket{0} \rightarrow \left( \ket{0} + \ket{1}\right)/\sqrt{2}$, 
$\ket{1} \rightarrow \left( \ket{0} - \ket{1}\right)/\sqrt{2}$ and vice-versa.  While the gates $P$ and $T$ 
rotate the phase of a single qubit by an angle of $\frac{\pi}{4}$ ($P$) and $\frac{\pi}{8}$ ($T$) respectfully.  
The set of gates ($H$, $P$ and $T$) can be used to construct an arbitrary single qubit gate, through 
the Solovay-Kitaev algorithm \cite{DN06} and any arbitrary single qubit gate + the CNOT can be used to construct 
\emph{any} $m$-qubit unitary.  

While there are many different choices for a universal set of quantum gates, we choose this particular set 
for one extremely important reason; error correction.  The necessity of QEC and fault-tolerant protocols 
places strong restrictions on the types of quantum gates that can used on encoded data.  The encoding structure 
for error corrected qubits does not allow for arbitrary gates to be applied.  For isolated qubits, the implementation 
of quantum gates is dependant on the physics of the system, i.e. performing a rotation on an individual 
electron by an arbitrary angle along some arbitrary axis is simply a function of how the electron is aligned 
within an external electromagnetic field.  In contrast, for a group of qubits that are used to encode a 
protected piece of quantum information, it is the symmetries of the underlying code that dictates which 
operations are valid.  For essentially all quantum codes that are experimentally useful, this set of valid 
gates is extremely small and the set given above is the most commonly used.  

\subsubsection{Error Correction}
The second component necessary for a realistic model is that of error correction.  As noted in the introduction, 
useful algorithms require component accuracies far beyond what is achievable experimentally.  Solving 
large factorisation problems or simulating processes in quantum chemistry would require, if 
error correction was not used, component failure rates at least $10^{-15}$.  The best experimental systems 
can routinely reach error rates of about $10^{-2}$, hence an improvement of 13 orders of magnitude would 
be needed before a non-error corrected computer could operate.  

Error correction solves this problem by encoding logical qubits into a group of multiple physical qubits 
using an appropriate code \cite{DMN09}.  By repeated encoding, and by performing operations 
in such a way that physical errors do not cascade (i.e. a single error gets copied to a 
large number of errors), a process known as Fault-tolerance, the logical information can be 
protected to an arbitrary level at the expense of more physical qubits and longer processing times.  

This process does not work with arbitrarily bad components, each physical device must have a minimum 
level of accuracy such that the additional operations for QEC do not introduce more errors than the code 
is designed to correct.  This minimum level of accuracy is refereed to as the fault-tolerant threshold and 
represents the minimum \emph{physical} error rate tolerable, per qubit, per time step such that error 
correction will be effective and arbitrary computation possible.  The goal of error correction and architectural 
design is to find and integrate QEC codes that have a high threshold and can be deployed on a physical 
system which may have many constraints associated with qubit placement, interactions and transport.  

\section{Topological Computation (TQC)}  Most modern quantum architectures are based upon the model of TQC \cite{DFSG08, MLFY10, JMFMKLY10}, as this method for computation has QEC integrated by construction. TQC is the preferred method for three primary reasons. 1) It has one of the highest fault-tolerant thresholds of any method of QEC.   2) It is a \textit{local} model of computation, i.e. individual physical qubits in the computer only have to interact with their neighbours. 3) The quantum hardware is only used to prepare a large three-dimensional lattice of connected qubits (the topological lattice), algorithmic implementation is a function of how the lattice is consumed rather than how it is created, i.e. it is a measurement based model. Therefore the TQC model is a software driven method of computation. 
\begin{figure}[t!]
	\centering
	\includegraphics[scale=0.2]{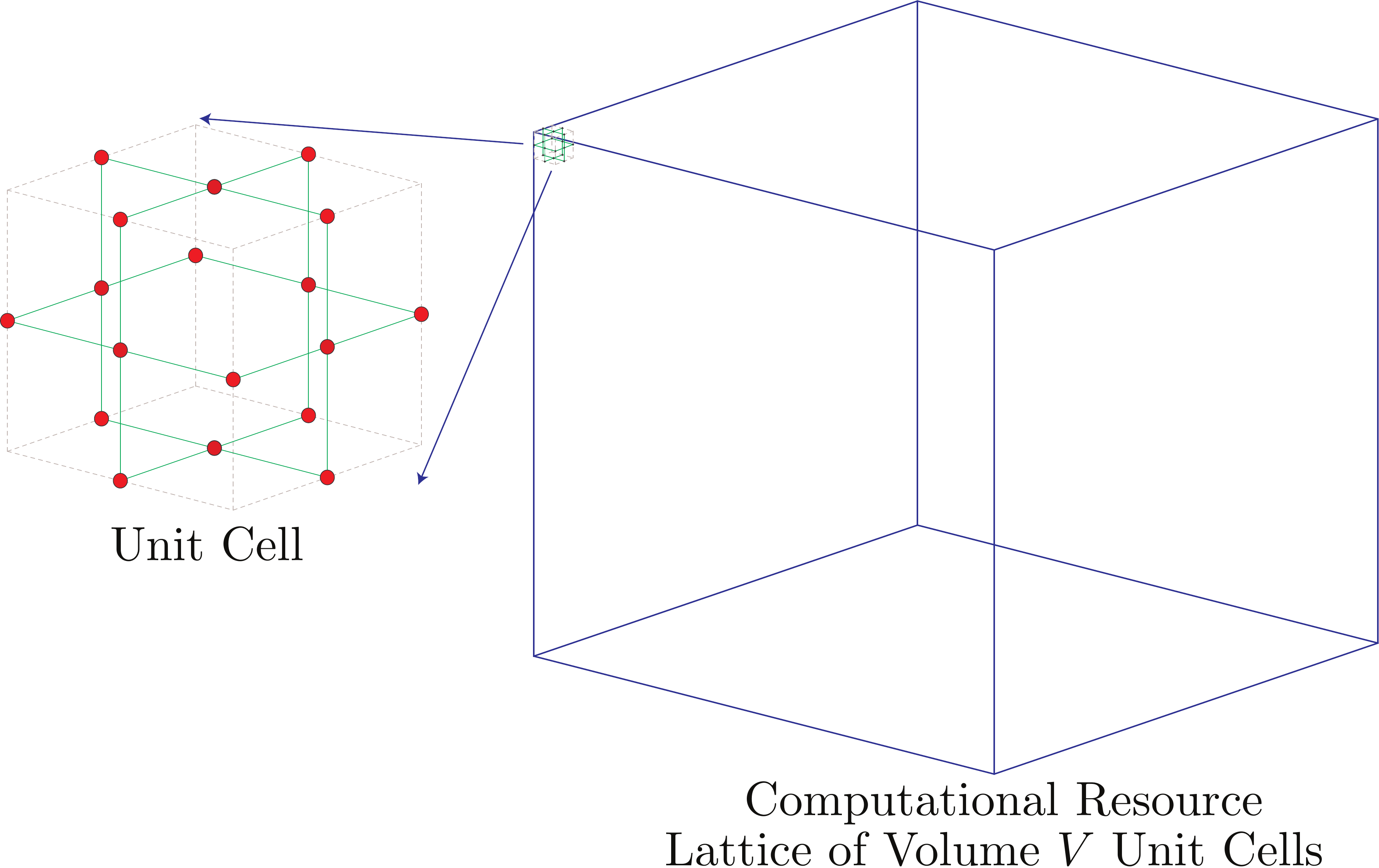}
	\caption{Structure of the quantum state needed for TQC.  The quantum hardware produces a large 
	lattice of qubits that are connected (entangled) with four of their nearest neighbours. This massive 3D 
	lattice provides the workbench where computation proceeds via the strategic consumption of individual 
	qubits.  The large 3D volume is built from unit cells of 18 qubits, the connections form a regular unit 
	cell that extends in all three dimensions to fill the volume.}
	\label{fig:cell}
\end{figure}

\begin{figure}[t!]
	\centering
	\includegraphics[scale=0.15]{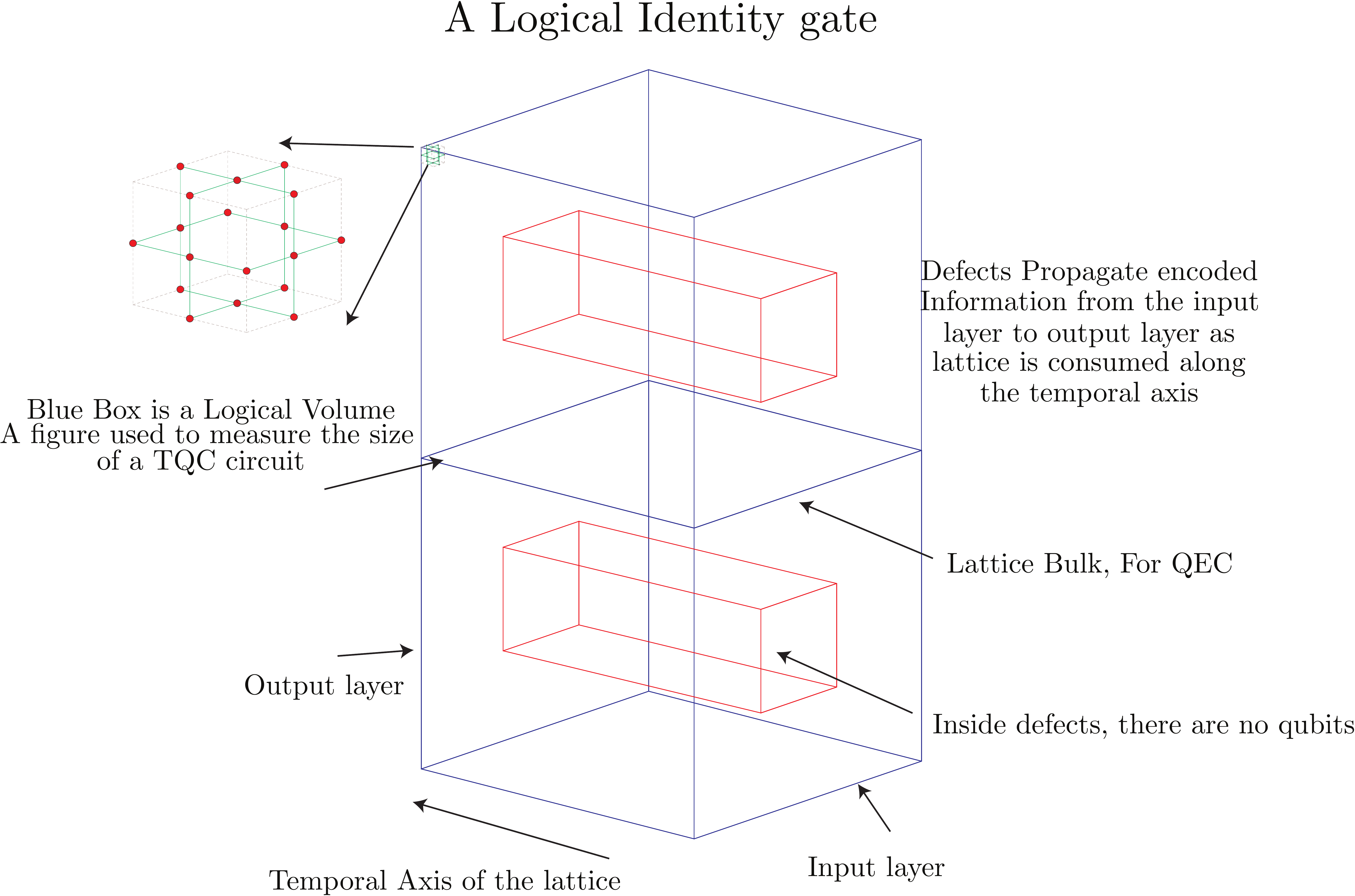}
	\caption{\textit{A Logical Identity operation}.  One axis of the 3D lattice is defined as the temporal 
	axis for computation.  An encoded qubit is defined via a pair of defects (holes in the lattice).  These 
	defects propagate encoded information from the input layer 
	(consumed at some time $t$) to the output layer (consumed at time $t' > t$).  Information is protected from 
	errors by creating large defects that are significantly separated by the lattice bulk.  A logical cell 
	can be defined (the outer boxes surrounding each defect) allowing us to measure the size of a 
	quantum circuit in terms of the spatial/temporal resources produced by the hardware.}
	\label{fig:logicalVol}
\end{figure}

The specifics of how TQC works can be found in the following References \cite{RHG07,FG08}, 
we will provide a more conceptual summary of the basis principals surrounding TQC.  The quantum hardware prepares a massive 3D lattice of qubits that are all connected (entangled) to form a single enormous quantum state.  The unit cell of this lattice is shown in Fig.~\ref{fig:cell}.  Before computation proceeds, the hardware 
simply prepares a "clean" lattice.  i.e. it is a single, \emph{unique} quantum state and contains no encoded 
information.  Before computation begins, one dimension of this lattice is identified with the temporal axis of 
computation.  Information is propagated along this temporal axis and gate operations are arranged in 
this direction, reflecting the underlying algorithm.  

Logical information is introduced and error protected by deliberately creating holes in this lattice, 
called defects.  Shown in Fig.~\ref{fig:logicalVol} is an example of a logically encoded qubit, undergoing 
a simple identity operation, defined via a pair of defects.  
For the identity gate, information is propagated from an input layer to an output layer 
(at a later time step) along the defect. The defect is created and 
propagated along the temporal axis by simply removing the physical qubits internal to its boundary.  In Fig.~\ref{fig:logicalVol}, the defects are the red rectangular structures and all physical qubits internal to these 
structures are removed from the lattice.  This removal can be achieved one of two ways.  
Either the physical qubits are physically discarded from the lattice or these qubits can be measured in the $\ket{0}$, 
$\ket{1}$ basis.  Measurement in this basis disconnects all the bonds from the respective qubit and has the 
same effect as simply removing them.  

The defect is surrounded by a region of the lattice which is the bulk.  The bulk is responsible for 
the error correction.  Logical information is corrupted in this model if a chain of \emph{physical} 
qubits that connect one defect to another or create a closed loop encircling a defect experience errors.  
Therefore, if a defect has a large cross-section and is surrounded by a large "buffer" of the bulk, the information is heavily protected.  Increasing the cross sectional size of the defect or the size of the bulk linearly 
reduces the error rate of the encoded information by approximately an exponential factor.  

As one axis of the lattice represents the temporal direction of computation, the encoded information propagates 
from an input layer to the output layer.  
The purpose of computation with this model is to, in a controlled manner, manipulate the shape and movement of the defects within a large lattice produced by the hardware. 

\subsection{Other gates}
The previous section illustrated one gate that can be implemented in the TQC model, here we will examine 
the other operations that can be implemented directly.  This section will focus on the geometric 
structures that represent certain operations, the details of why these structures realise such gates can be 
found in \cite{RHG07,FG08}
\subsubsection{Measurement and Initialization} 
\begin{figure}[t!]
	\centering
	\includegraphics[scale=0.22]{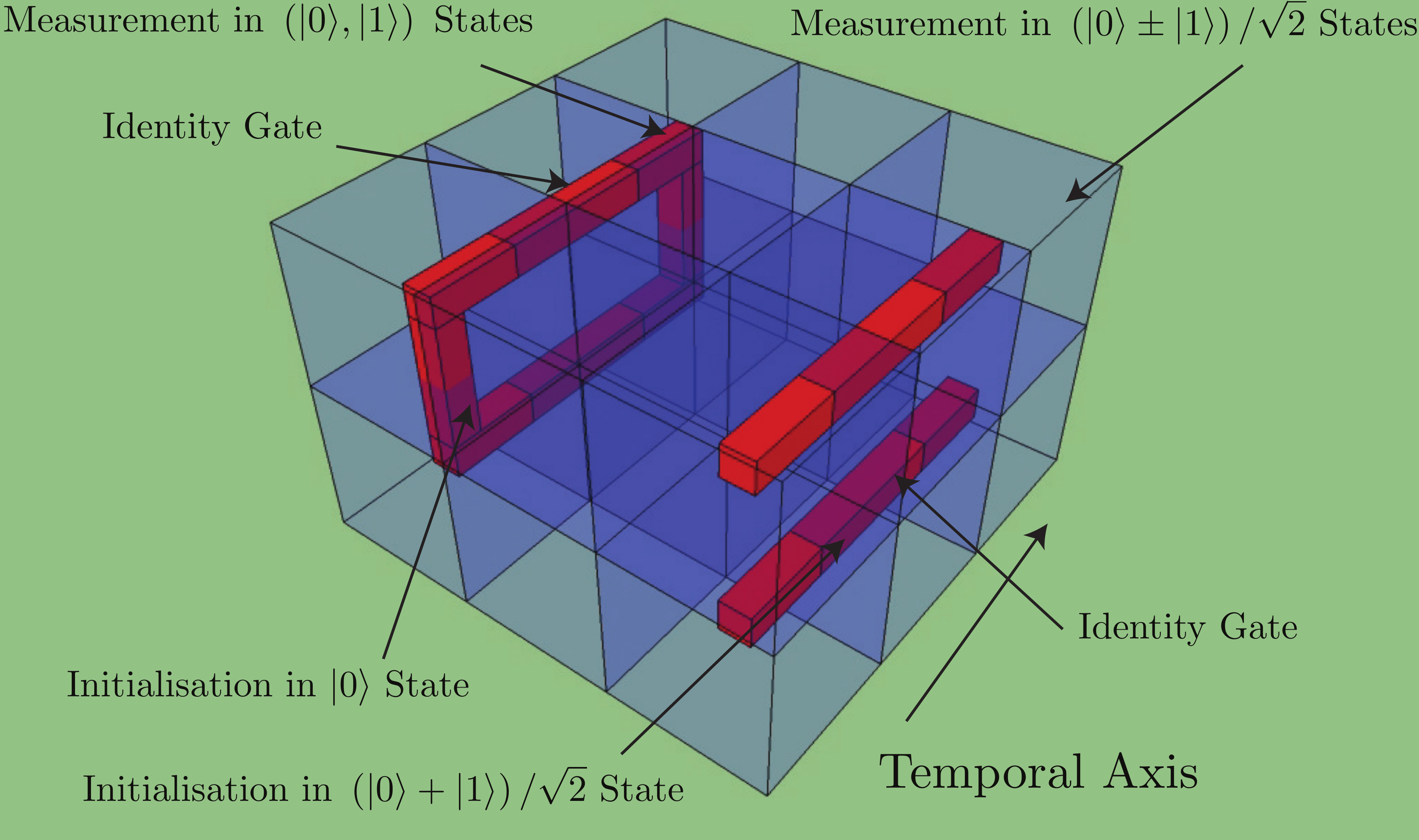}
	\caption{Diagrams of initialisation and measurement in the TQC model, illustrating the two types of 
	initialisation states and measurement bases allowed by the topological code.}
	\label{fig:initial}
\end{figure}

The lattice allows for only a restricted set of states that can be initialised directly and a restricted set 
of possible measurements.  Only the states $\ket{0}$  
and $\left(\ket{0}+\ket{1}\right)/\sqrt{2}$ can be initialised fault-tolerantly. Fault-tolerant 
measurements can only be made of the states ($\ket{0}$, $\ket{1}$) and $\left(\ket{0}\pm \ket{1}\right)/\sqrt{2}$.   The geometric structures for these operations 
are illustrated in Fig. \ref{fig:initial}.

We have shown two sets of structures.  The one on the left illustrates the initialisation of an encoded qubit in the 
$\ket{0}$ state, a horseshoe structure that is created at a certain point as the lattice is consumed, an identity
gate by maintaining the defects in straight lines and a measurement in the ($\ket{0}$, $\ket{1}$) basis, which 
is the time reversed horseshoe structure corresponding to initialisation.  Each of these steps requires 
a logical volume of two \emph{logical} cells.  The total volume for this small circuit is therefore six.  
The second structure on the right illustrates the same, but this time we initialise the encoded qubit in the 
$\left(\ket{0}+\ket{1}\right)/\sqrt{2}$ state and measured in the $\left(\ket{0} \pm \ket{1}\right)/\sqrt{2}$ basis.  
Again this circuit requires a logical volume of six.  When we initialise or measure the encoded qubits 
we are again simply choosing to begin removing qubits from the lattice at the points defined by the 
defects.

\subsection{Primal and Dual defects}
Before we discuss a more complicated gate, we first need to introduce the idea of primal and 
dual defects.  The structure of the lattice imbeds two self similar lattices.  Fig. \ref{fig:dual} 
illustrates. 
\begin{figure}[t!]
	\centering
	\includegraphics[scale=0.45]{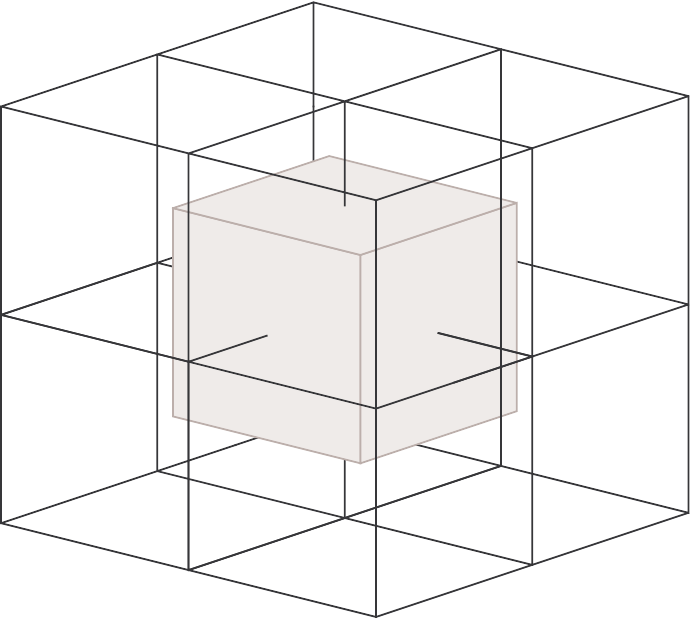}
	\caption{\textit{The primal and dual spaces of the lattice.}  If you combine eight unit cells of the 
	lattice, then at the intersection of those eight cells, you find an identical unit cell.  Defects can be 
	created by removing qubits from the faces of primal cells (giving arise to primal qubits) or 
	dual cells (giving dual defects).  Qubits can be encoded using either type.}
	\label{fig:dual}
\end{figure}
By combining eight cells of the lattice an identical unit cell structure exists at the intersection of these 
eight cells.  This is a unit cell in the dual lattice.  The dual lattice is offset from the primal lattice by 
half a unit cell along all three spatial axes.  As defects can be defined via the removal of selected 
face qubits from primal lattices, we can do the same for face qubits in the dual cells.  This then defines 
a dual type defect.  Dual defects behave identically to primal defects except that the initialisation and 
measurement structures shown in Fig. \ref{fig:initial} are reversed (i.e. the horseshoe structures 
represent initialisation in $\left(\ket{0}+\ket{1}\right)/\sqrt{2}$ and measurement in the 
$\left(\ket{0}\pm\ket{1}\right)/\sqrt{2}$ basis rather than $\ket{0}$, $\ket{1}$).  

\subsubsection{CNOT gate}
The main reason for introducing the concept of primal and dual defects is to explain the structure of the 
logical CNOT gate.  The logical CNOT gate is achieved using a concept known as braiding.  This is where 
defects are moved around each other.  For this gate to be effective, it \emph{must be performed using 
defects of opposite} type.  If braiding is performed using defects of the same type, no interaction 
will take place.  
\begin{figure}[ht!]
	\centering
	\includegraphics[scale=0.20]{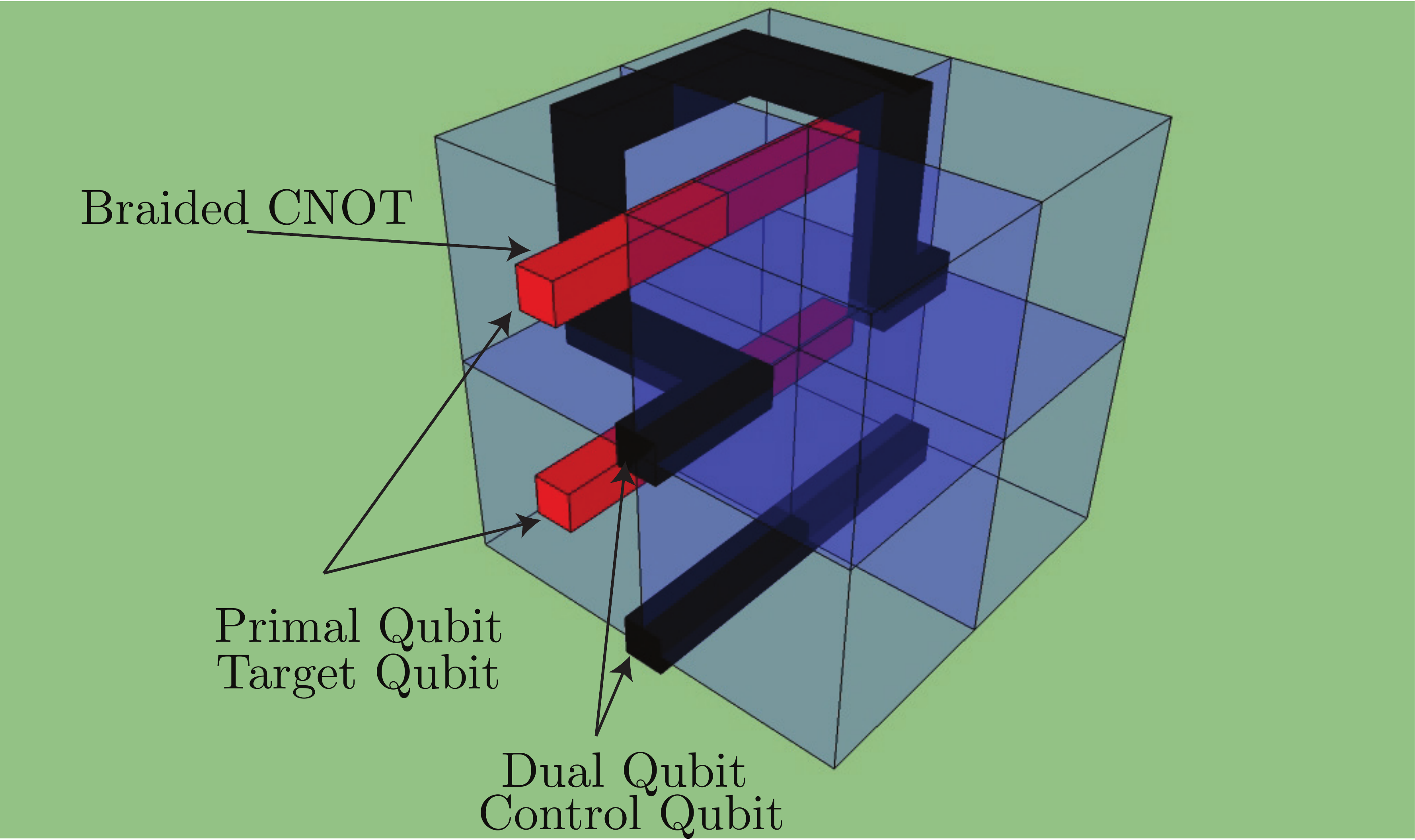}
	\caption{\textit{A braided CNOT gate.}  A CNOT interaction takes place between two defects of opposite 
	type.  One of the two defects of the encoded qubit is manipulated such that it braids around one 
	defect of the other encoded qubit as the lattice is consumed.}
	\label{fig:dual}
\end{figure}

Illustrated in Fig. \ref{fig:dual} is a CNOT performed between a primal defect (red) and dual defect (black). 
It should be noted that the dual qubit is \emph{always the control qubit} for the interaction.  As with the other gates 
illustrated, movement of the defects occurs as the lattice is consumed and is defined by which physical qubits 
are removed from the lattice.

This CNOT is the main interaction gate that is utilised in the topological model.  However it can 
only occur between defects of opposite type.  This is not desirable for large scale computation as many 
different pairs of encoded qubits need to be interacted during an algorithm and we cannot simply partition all 
of them into sets of primal and dual.  We ideally want to perform a CNOT between defects of \emph{the same type}.  

\subsection{Performing a CNOT between two primal encoded qubits}
\begin{figure}[t!]
	\centering
	\includegraphics[scale=0.33]{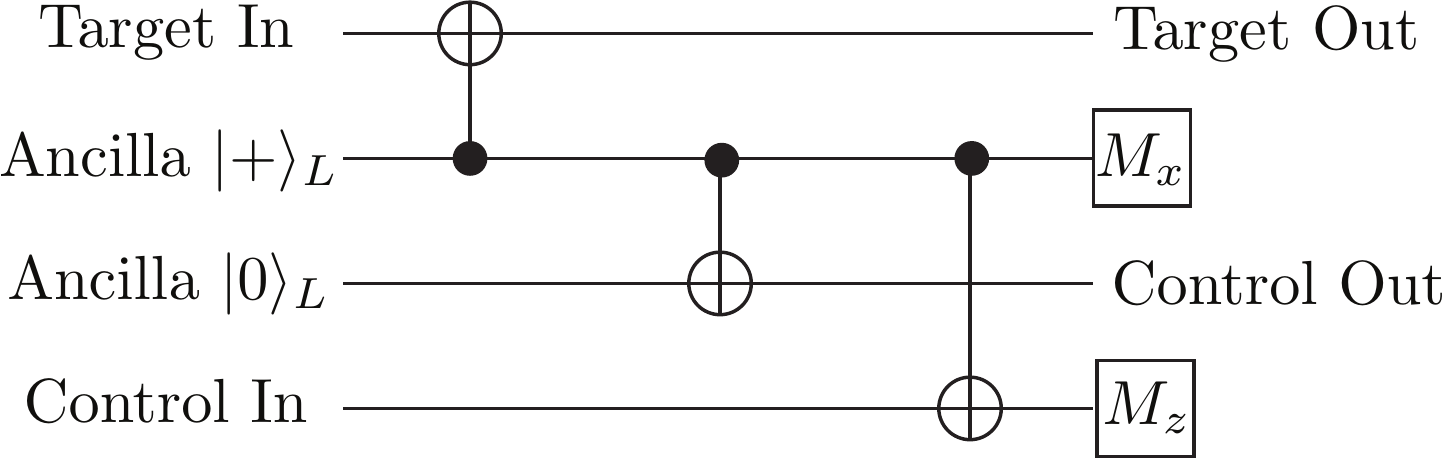}
	\caption{In this circuit identity a CNOT is constructed using four qubits and three CNOT gates.  
	Here the control for each CNOT is the same qubit.  Therefore, if this qubit is a dual qubit, then each of 
	the other three can be primal.  This allows a CNOT between two primal encoded qubits.}
	\label{fig:identity}
\end{figure}

Being able to perform a CNOT between two primal encoded qubits requires us to consider the following 
circuit identity [Fig. \ref{fig:identity}].  
This identity simply allows for a CNOT gate between the control and target input by introducing two extra 
qubits (initialised into the $\ket{0}$ and $\ket{+} = \left(\ket{0}+\ket{1}\right)/\sqrt{2}$ states), performing 
three CNOTS and measuring out two of the qubits.  The reason this identity is useful is because one of 
the ancilla qubit act as control for all three gates.  Therefore, if this qubit is an encoded dual qubit, we 
can realise a CNOT between two primal encoded qubits.  

\begin{figure}[t!]
	\centering
	\includegraphics[scale=0.21]{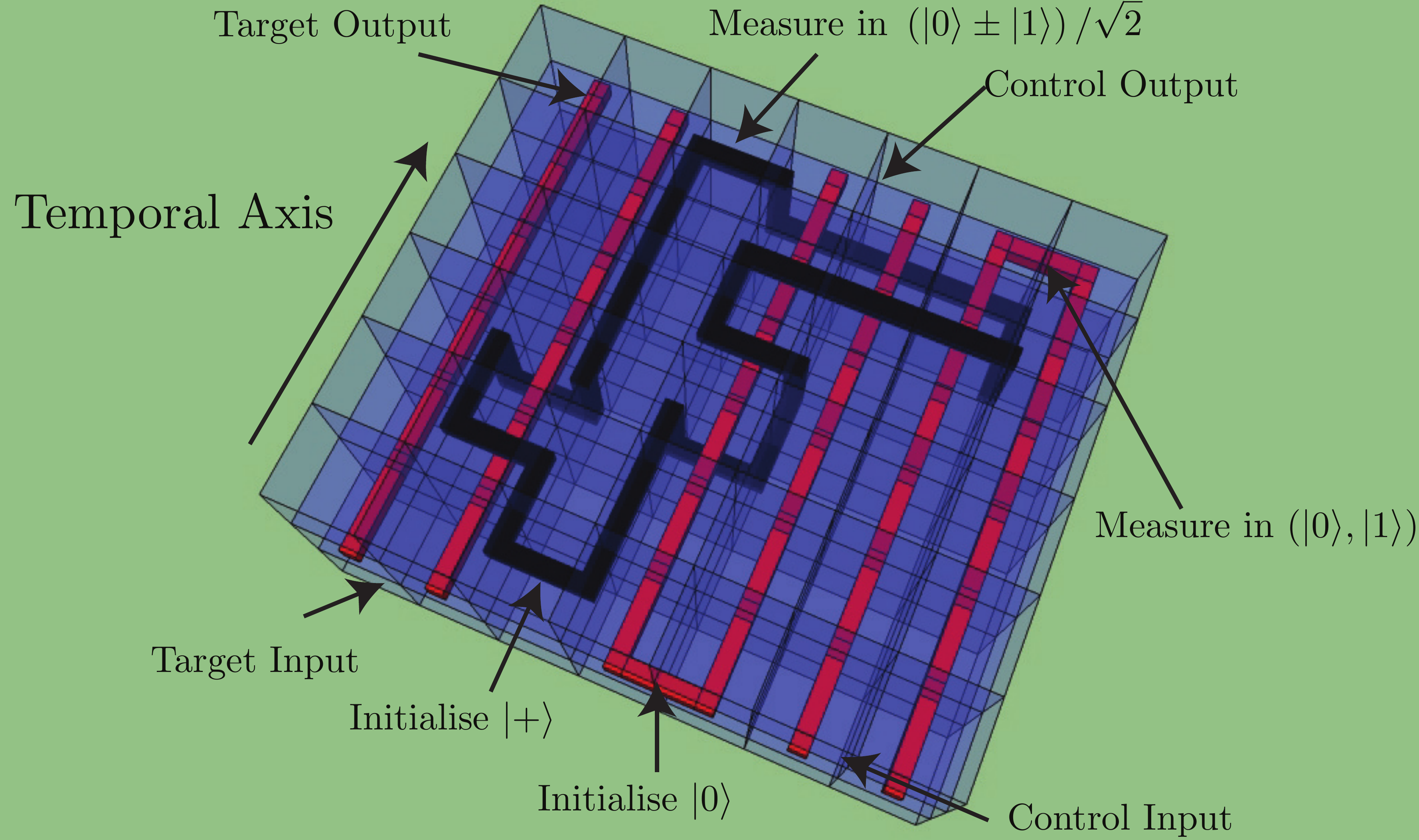}
	\caption{\textit{CNOT between to primal qubits}. Combining the braiding structure in Fig. \ref{fig:dual} with 
	the circuit of Fig. \ref{fig:identity} allows us to generate the required operation.  The volume of lattice 
	occupied by this gate is quite large.}
	\label{fig:CNOT2}
\end{figure}

This circuit structure can be mapped directly to a braiding pattern for topological computation, illustrated 
in Fig. \ref{fig:CNOT2}.  This modified CNOT is constructed by using the structure in Fig. \ref{fig:dual} and 
the circuit of Fig. \ref{fig:identity}.  

\section{Compactifying circuits}
The topological circuit of Fig. \ref{fig:CNOT2} looks to be very inefficient in terms of lattice volume.  The CNOT 
of Fig. \ref{fig:identity} occupies a volume of eight cells, but the CNOT between two primal encoded qubits 
requires a volume of 126 logical cells.  This is where the idea of compactifying circuits can be introduced.  

What follows is essentially the essence of this introduction and represents the primary goal for a compiler 
for topological computation.  Defects are allowed to be manipulated in various ways provided the underlying 
topology of the circuit is maintained \footnote{Follow up papers will introduce an array of legal 
moves and how they can be used to compactly circuits}.  In the case of the CNOT, this simply requires us to maintain the manner 
in which each individual defect strand is braided with the others.  

\begin{figure}[t!]
	\centering
	\includegraphics[scale=0.11]{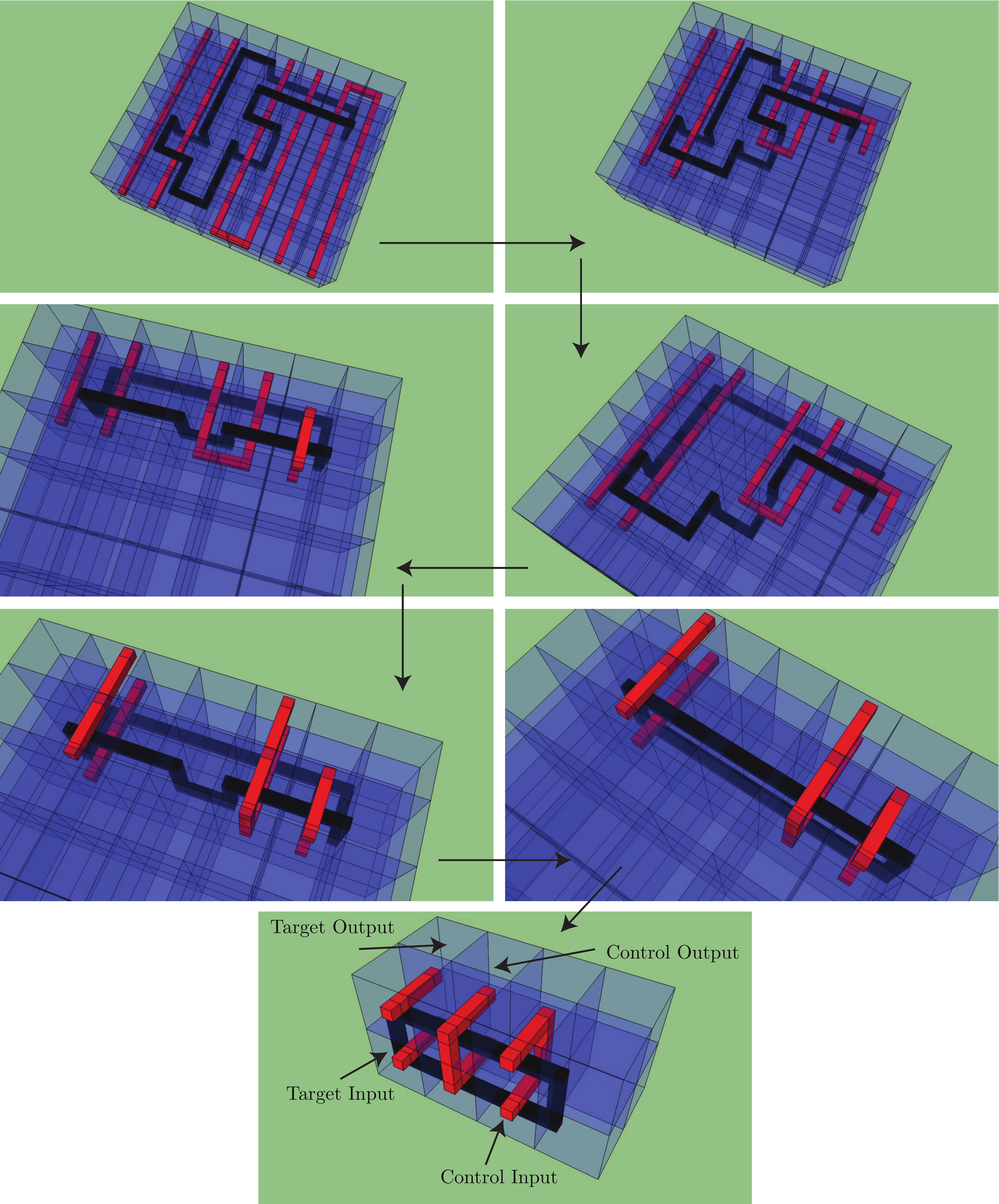}
	\caption{\textit{Compactifying the CNOT}.  A series of continuous 
	deformations can be used to reduce the size of the circuit.  Moves are 
	valid if they preserve the topology (i.e. the manner in which defects are braided together)}
	\label{fig:reduce}
\end{figure}

As the CNOT is a relatively simple example, we can illustrate explicitly some of the movements that can be 
made to the defect structure that reduces the total lattice volume needed to implement the gate.  This sequence is
 illustrated in Fig. \ref{fig:reduce}
 The final, reduced version of the circuit has a volume of 16, representing a reduction in lattice volume of a 
 factor of 7.9.  This represents a significant saving of hardware resources as each logical cell of the lattice may 
 contain many thousands of physical qubits.  Additionally the number of cells along the temporal axis of the 
 lattice has been reduced from 6 to 2, this increases the speed of the logical gate by a factor of three.  

\section{An example of a Larger circuit}
\begin{figure}[ht!]
	\centering
	\includegraphics[scale=0.16]{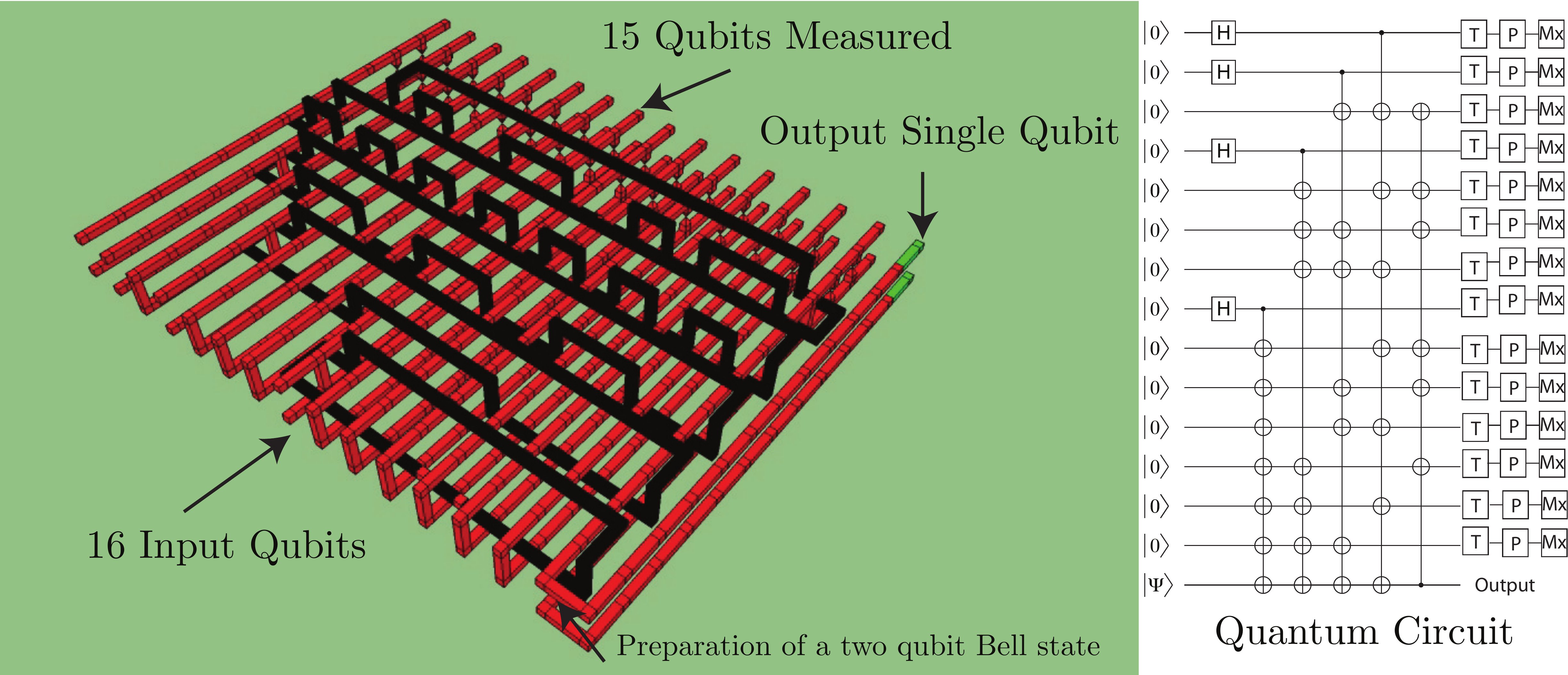}
	\caption{\textit{Purification braiding circuit for the state $\left(\ket{0}+e^{i\frac{\pi}{4}}\right)/\sqrt{2}$}. 
	This is the most used single circuit in a quantum algorithm. The actual quantum circuit is also shown on the 
	right.  The circuit requires 16 qubits, and five sets of multi-target CNOT gates.  
	The small pyramid structures that are slightly visible towards the output are known as injection points, which 
	introduce high error states that are purified by the circuit \cite{RHG07,FG08} (these 
	circuits are used to implement $H$, $P$ and $T$ gates).}
	\label{fig:Tstate1}
\end{figure}

Finally, as an example, we illustrate the structure for a larger quantum circuit.  Fig. \ref{fig:Tstate} is the 
quantum circuit for a process known as state distillation \cite{BK05+}, with the 
braid pattern shown in Fig. \ref{fig:Tstate1}.  This circuit is required for the fault-tolerant 
application of the $T$ gate [Eq. \ref{Eq:Tstate}], which by far is the most used circauit in a large scale 
quantum algorithm.  By some estimates \cite{JMFMKLY10} this circuit can represent above 80\% of all operations within 
a large quantum algorithm.  Only the CNOT, identity, initialisation and measurement can 
be applied directly to the topological lattice.  The other three gates ($H$, $P$ and $T$) forming 
a universal gate set are applied through teleportation operations and distillation protocols, ultimately 
constructed from large CNOT networks \cite{FG08,PDNP12}.

The circuit is designed to increase the purity of a single qubit encoded state 
$\ket{Y} = \left(\ket{0} + e^{i\frac{\pi}{4}}\ket{1}\right)/\sqrt{2}$ from 15 "dirty" copies of the same state.  This is to allow us to 
perform certain very low error quantum gates than cannot be directly implemented in the TQC model \cite{RHG07,FG08}.  This 
specific circuit has a volume in the topological lattice of 384.  Reducing the volume required for this circuit 
will significantly decrease the qubit/time resources required for any large scale algorithm.  Hence for 
any optimisation process, this circuit should be considered first \cite{FD12}.

\section{Conclusion}
In this paper we have introduced the concept of topological quantum computation and described the problem 
of optimisation for large quantum circuits.  This introduction was done in a very conceptual manner.  The goal of any successful optimisation program is 
to compact a large quantum algorithm consisting of many components into a 
3D geometric braid diagram that occupies the smallest possible volume of the lattice produced by the 
hardware.  Future papers will explain the rules of how encoded defects can be manipulated.  

This paper is intended as a very preliminary explanation of the general problem.  Those fluent in the language 
of quantum information science can read the associated papers to gain a better understanding of the 
issues related to optimisation.  

This field, which we are dubbing "Quantum Informatics" has just begun, and hopefully in the near future many 
in the field of classical computer science will examine the issues related to programming a 
topological quantum computer and help us develop appropriate software packages to design and 
optimise massive topological quantum circuits.

\section*{Acknowledgements}
This work is supported by the Quantum Cybernetics (MEXT) and FIRST projects, Japan.

% trigger a \newpage just before the given reference

%\IEEEtriggeratref{8}
% The "triggered" command can be changed if desired:
%\IEEEtriggercmd{\enlargethispage{-5in}}

\bibliography{paper}
\bibliographystyle{IEEEtran}

\end{document}